\documentstyle[twoside,multicol,psfig]{article}
\catcode`\@=11
\long\def\@makefntext#1{
\protect\noindent \hbox to 3.2pt {\hskip-.9pt
$^{{\footnotesize\@thefnmark}}$\hfil}#1\hfill}      

\def\@makefnmark{\hbox to 0pt{$^{\@thefnmark}$\hss}}    

\def\ps@myheadings{%
    \let\@oddfoot\@empty\let\@evenfoot\@empty
    \def\@evenhead{\footnotesize\it\leftmark\hfil}
    \def\@oddhead{\hfil{\footnotesize\it\rightmark}}
    \let\@mkboth\@gobbletwo
    \let\sectionmark\@gobble
    \let\subsectionmark\@gobble
    }

\oddsidemargin=\evensidemargin
\newcounter{sectionc}\newcounter{subsectionc}\newcounter{subsubsectionc}
\renewcommand{\section}[1] {\vspace{14pt}\addtocounter{sectionc}{1}
\setcounter{subsectionc}{0}\setcounter{subsubsectionc}{0}\noindent
    {\bf\thesectionc. #1}\par\vspace{8pt}}
\renewcommand{\subsection}[1] {\vspace{14pt}\addtocounter{subsectionc}{1}
   \setcounter{subsubsectionc}{0}\noindent
   {\bf\thesectionc.\thesubsectionc. {\kern1pt \bfit #1}}\par\vspace{8pt}}
\renewcommand{\subsubsection}[1] {\vspace{14pt}
    \addtocounter{subsubsectionc}{1}
    \noindent{\thesectionc.\thesubsectionc.\thesubsubsectionc.
    {\kern1pt \it #1}}\par\vspace{8pt}}

\newcounter{appendixc}
\newcounter{subappendixc}[appendixc]
\newcounter{subsubappendixc}[subappendixc]
\renewcommand{\thesubappendixc}{\Alph{appendixc}.\arabic{subappendixc}}
\renewcommand{\thesubsubappendixc}
    {\Alph{appendixc}.\arabic{subappendixc}.\arabic{subsubappendixc}}

\renewcommand{\appendix}[1] {\vspace{14pt}
        \refstepcounter{appendixc}
        \setcounter{figure}{0}
        \setcounter{table}{0}
        \setcounter{lemma}{0}
        \setcounter{theorem}{0}
        \setcounter{corollary}{0}
        \setcounter{definition}{0}
        \setcounter{equation}{0}
        \renewcommand{\thefigure}{\Alph{appendixc}.\arabic{figure}}
        \renewcommand{\thetable}{\Alph{appendixc}.\arabic{table}}
        \renewcommand{\theappendixc}{\Alph{appendixc}}
        \renewcommand{\thelemma}{\Alph{appendixc}.\arabic{lemma}}
        \renewcommand{\thetheorem}{\Alph{appendixc}.\arabic{theorem}}
        \renewcommand{\thedefinition}{\Alph{appendixc}.\arabic{definition}}
        \renewcommand{\thecorollary}{\Alph{appendixc}.\arabic{corollary}}
        \renewcommand{\theequation}{\Alph{appendixc}.\arabic{equation}}
        \noindent{\bf Appendix \theappendixc #1}\par\vspace{5pt}}
\newcommand{\subappendix}[1] {\vspace{14pt}
        \refstepcounter{subappendixc}
        \noindent{\bf Appendix \thesubappendixc. {\kern1pt \bfit #1}}
    \par\vspace{8pt}}
\newcommand{\subsubappendix}[1] {\vspace{14pt}
        \refstepcounter{subsubappendixc}
        \noindent{\rm Appendix \thesubsubappendixc. {\kern1pt \it #1}}
    \par\vspace{8pt}}

\newcommand{\textlineskip}{\baselineskip=13pt}
\newcommand{\smalllineskip}{\baselineskip=10pt}

\newcommand{\copyrightheading}[1]
    {\vspace*{-2.5cm}\smalllineskip{\flushleft
    {\footnotesize International Journal of Neural Systems, #1}\\
    {\footnotesize \copyright\, World Scientific Publishing
     Company}\\
     }}

\newcommand{\publisher}[2]{{\begin{center}\tenrm\baselineskip=12pt
    Received #1\\
    Revised #2
    \end{center}
    }}

\def\abstracts#1#2{{
    \centering{\begin{minipage}{5.8in}\small\baselineskip=11pt
    \parindent=0pc #1\par
    \parindent=2pc #2
    \end{minipage}}\par}}

\renewenvironment{thebibliography}[1]       
    {\small\baselineskip=11pt
     \frenchspacing
     \begin{list}{\arabic{enumi}.}
        {\usecounter{enumi}\setlength{\parsep}{0pt}
     \setlength{\leftmargin 12.7pt}{\rightmargin 0pt}
         \setlength{\itemsep}{0pt} \settowidth
    {\labelwidth}{#1.}\sloppy}}{\end{list}}

\newcounter{itemlistc}
\newcounter{romanlistc}
\newcounter{alphlistc}
\newcounter{arabiclistc}

\newenvironment{romanlist}
    {\setcounter{romanlistc}{0}
     \begin{list}{$($\roman{romanlistc}$)$}
    {\usecounter{romanlistc}
     \setlength{\parsep}{0pt}
     \setlength{\itemsep}{0pt}}}{\end{list}}

\newcommand{\fcaption}[1]{
        \refstepcounter{figure}
        \setbox\@tempboxa = \hbox{\small Fig.~\thefigure. #1}
        \ifdim \wd\@tempboxa > 5in
           {\begin{center}
        \parbox{5in}{\small\baselineskip=11pt Fig.~\thefigure. #1}
            \end{center}}
        \else
             {\begin{center}
             {\small Fig.~\thefigure. #1}
              \end{center}}
        \fi}

\newcommand{\tcap}[1]{
        \refstepcounter{table}
        \setbox\@tempboxa = \hbox{\small Table~\thetable. #1}
        \ifdim \wd\@tempboxa > 3.15in
           {\begin{center}
        \parbox{3.15in}{\small\smalllineskip
        Table~\thetable. #1}
            \end{center}}
        \else
             {\begin{center}
             {\eightpoint Table~\thetable. #1}
              \end{center}}
        \fi}

\def\@citex[#1]#2{\if@filesw\immediate\write\@auxout
    {\string\citation{#2}}\fi
\def\@citea{}\@cite{\@for\@citeb:=#2\do
    {\@citea\def\@citea{,}\@ifundefined
    {b@\@citeb}{{\bf ?}\@warning
    {Citation `\@citeb' on page \thepage \space undefined}}
    {\csname b@\@citeb\endcsname}}}{#1}}

\newif\if@cghi
\def\cite{\@cghitrue\@ifnextchar [{\@tempswatrue
    \@citex}{\@tempswafalse\@citex[]}}
\def\citelow{\@cghifalse\@ifnextchar [{\@tempswatrue
    \@citex}{\@tempswafalse\@citex[]}}
\def\@cite#1#2{{$\null^{#1}$\if@tempswa\typeout
    {IJCGA warning: optional citation argument
    ignored: `#2'} \fi}}

\def\pmb#1{\setbox0=\hbox{#1}
    \kern-.025em\copy0\kern-\wd0
    \kern.05em\copy0\kern-\wd0
    \kern-.025em\raise.0433em\box0}

\def\fnt#1#2{\footnotetext{\kern-.3em
    {$^{\mbox{\scriptsize #1}}$}{#2}}}

\font\tenrm=cmr10
\font\tenit=cmti10

\font\bfit=cmbxti10 at 10pt

\font\eightit=cmti8

\def\itlatex{\tenit L\kern-.30em\raise.4ex\hbox{\eightit A}\kern-.14em
T\kern-.1667em\lower.7ex\hbox{E}\kern-.125em X}

\def\bsc{{\sc a\kern-7pt\sc a}}
\def\bflatex{\bf L\kern-.30em\raise.3ex\hbox{\bsc}\kern-.18em
T\kern-.1667em\lower.7ex\hbox{E}\kern-.125em X}

\def\qed{\hbox{${\vcenter{\vbox{            
   \hrule height 0.4pt\hbox{\vrule width 0.4pt height 6pt
   \kern5pt\vrule width 0.4pt}\hrule height 0.4pt}}}$}}

\begin{document}
\setlength{\textheight}{8.78truein}     

\thispagestyle{empty}

\markboth{Emergence of scale-free properties in Hebbian networks}
{Emergence of scale-free properties in Hebbian networks}

\textlineskip
\setcounter{page}{1}

\copyrightheading{Vol.~0, No.~0 (April, 2000) 00--00}

\vspace*{1.05truein}


\centerline{\large\bf Emergence of scale-free properties in
Hebbian networks}

\vspace*{0.45truein} \centerline{G\'abor Szirtes}
\vspace*{0.0215truein} \centerline{\it E-mail:
guminyul@ludens.elte.hu} \vspace*{14pt}\centerline{Zsolt Palotai}
\vspace*{0.0215truein}  \centerline{\it E-mail: pz120@hszk.bme.hu}
\vspace*{14pt} \centerline{Andr\'as L{\H
o}rincz\footnote{corresponding author}} \centerline{\it E-mail:
lorincz@inf.elte.hu} \vspace*{0.0215truein} \centerline{\it
Department of Information Systems, E\"otv\"os Lor\'and University
} \baselineskip=11pt \centerline{\it P\'azm\'any P\'eter
s\'et\'any 1/C, Budapest, Hungary H-1117, Phone: +36 1 381 2143,
Fax: +36 1 381 2140}

\vspace*{0.3truein}
\publisher{~(to be inserted}{~by Publisher)}

\vspace*{0.29truein} \abstracts{The fundamental `plasticity' of
the nervous system (i.e high adaptability at different structural
levels) is primarily based on Hebbian learning mechanisms that
modify the synaptic connections. The modifications rely on neural
activity and assign a special dynamic behavior to the neural
networks. Another striking feature of the nervous system is that
spike based information transmission, which is supposed to be
robust against noise, is noisy in itself: the variance of the
spiking of the individual neurons is surprisingly large which may
deteriorate the adequate functioning of the Hebbian mechanisms. In
this paper we focus on networks in which Hebbian-like adaptation
is induced only by external random noise and study spike-timing
dependent synaptic plasticity. We show that such `HebbNets' are
able to develop a broad range of network structures, including
scale-free small-world networks. The development of such network
structures may provide an explanation of the role of noise and its
interplay with Hebbian plasticity. We also argue that this model
can be seen as a unification of the famous Watts-Strogatz and
preferential attachment models of small-world nets.}{Keywords:
small world, Hebbian learning, central nervous system, scale-free
network}

\vspace*{10pt}\textlineskip

\begin{multicols}{2}
\section{Introduction}
\label{s:intro}

In the last few years spike-timing dependent synaptic plasticity
(STDP) (see e.g. (Ref.~1) and references therein), which is an
extension of the classical Hebbian learning mechanism, has been
the subject of intensive research. Recent experiments \cite{2,3,4}
(for a review, see, e.g. (Ref. 5)) revealed that exact timing and
temporal dynamics of the neural activities play a crucial role in
forming the neuronal base of plasticity. While it is still an open
question, whether the rate of spikes (that is temporal or
population averaged spike count) or the exact time pattern of the
spikes carries the information, it is broadly accepted in the
machine learning literature\cite{6,7,8} and is strongly supported
in neuronal modelling\cite{9} that spike based encoding can be
efficient in compression, allows for sparse representation, low
energy consumption and that it can be robust against noise. The
last property seems to be indispensable knowing the stochastic
behavior of the neurons and of the external environment. But if
noise should be suppressed, how come that a great part of the
signals propagating through several brain regions experienced in
different species (ranging from frogs to primates) is considered
to be internally generated noise\cite{10,11}? What can be the
reason for counteracting the perfect information processing and
transmission? One possible role of noise in the nervous system is
provided by the recognition that noise can enhance the response of
nonlinear systems to weak signals, via a mechanism known as
stochastic resonance (see, e.g., (Ref. 12)). However, noisy
functioning may have additional roles. For example, it has been
shown that synaptic background activity may promote distinguishing
very similar inputs\cite{13}. It has been also
demonstrated\cite{14} that strict conditions on stability of
Hebbian mechanisms can be released by introducing random external
noise instead of maintaining competition among neurons over the
input sets. In this paper we address the question whether noise
may have any impact on \textbf{structural} changes.

In the following, we examine what network structures may emerge in
a simplistic neural system by applying \textbf{pure} Hebbian
learning. From now on, this neuronal network model will be
referred as to \textit{HebbNet}.

\section{\label{s:model}Description of HebbNet}

 We assume that the network is \textit{sustained} by inputs with no
spatio-temporal structure; that is the input is random noise. Our
models consist of $N$ number of simplified integrate-and-fire-like
`neurons' or nodes. The dynamics of the internal activity is
written as
\begin{equation}
 \frac{\Delta a_i}{\Delta t} =\sum_{j} w_{ij}a_j^s+x_i^{(ext)},
 \label{e:int_and_fire}
\end{equation}
for $i=1,2,\ldots ,N$. (N was 200 in our simulations.) Variable
$x^{(ext)} \in {(0,1)}^N$ denotes the randomly generated input
from the environment, $a_i$ is the internal activity of neuron
$i$, $w_{ij}$ is $ij^{th}$ element of matrix $\mathbf{W}$, i.e.,
the connection strength from neuron $j$ to neuron $i$. If $\Delta
t = 1$ then we have a discrete-time network and each parameter has
a time index, or if $\Delta t$ is infinitesimally small then
Eq.~\ref{e:int_and_fire} becomes a set of coupled differential
equations. Neuron $j$ outputs a spike (neuron $j$ `fires') when
$a_j$ exceeds a certain level, the threshold parameter $\theta$.
Spiking means that the output of the neuron $a_j^s$ (superscript
$s$ stands for `spiking') is set to 1. After firing, $a_j$ is set
to zero at the next time step for the discrete-time network. For
the continuous version of Eq.~\ref{e:int_and_fire}, $a_j$ is set
to zero after a very small time interval. Amount of excitation
received by neuron $i$ from neuron $j$ is $w_{ij} a_j^s$. Equation
\ref{e:int_and_fire} describes the simplest form of
`integrate--and--fire' network models which is still plausible
from a neurobiological point of view. Note that if $\Delta t = 1$
and the threshold is set to zero (i.e., if a neuron receives any
excitation then it fires and is reset to zero) then
Eq.~\ref{e:int_and_fire} represents `binary neurons' without
temporal integration. This can be seen as the simplest model
within our framework. Also, if the threshold is kept and if $a_i$
is set to zero before each time step, irrespective if the $i^{th}$
neuron fires or not, then the original model of McCullough and
Pitts\cite{15} is recovered.

Beyond the local activity threshold, we also examined the effect
of global activity constraint: at each time instant, a given
percent of nodes was selected randomly in proportion to the
activity $a_i$ for all $i=1,2, \ldots , N$. These neurons fired at
that time instant. For these two cases, computer simulations
showed negligible differences.

Synaptic strengths are modified as follows:
\begin{equation}
 \frac{\Delta w_{ij}}{\Delta t}
=\sum_{(t_i,t_j)}K(t_j-t_i)a_i^{t_i,s}a_j^{t_j,s},
 \label{e:pot}
\end{equation}
where $K$ is a kernel function which defines the influence of the
temporal activity correlation on synaptic efficacy, $t_i, t_j$ the
spiking times of neuron $i$ and $j$, respectively and
$a_i^{t_i,s}$ is the firing activity of neuron $i$ at time $t_i$.
$\Delta w_{ij} / \Delta t$ may be taken over discrete or over
infinitesimally small time intervals. Possible kernels are
depicted in Fig.~\ref{f:kernelf}. The kernel is a function of the
time differences. Because, in our case, the input is noise with no
temporal correlation, only the ratio of the positive
(strengthening) and the negative (weakening) areas of the kernel
function ($r_{A^{+}/A^{-}}$) should count. Temporal grouping and
reshaping of the kernel would not modify our results as long as
the aforementioned ratio is kept constant and the input is pure
noise. For this special case, the difference between the two
kernel types of Fig.~\ref{f:kernelf} does not have much impact on
the temporal evolution of our model network. It should be noted
that including inputs with spatiotemporal structure and other
known details of synaptic plasticity mechanisms, this kernel shape
independence will not hold. Our only constraint on the kernel,
namely the constraint that $r_{A^{+}/A^{-}}<1$, is required to
constrain weights. This constraint redistributes weight strengths.
Empirical data indicate that indeed, there are mechanisms to
redistribute weight strengths; potentiation for weak synapses is
favored whereas strong synapses tend to be depressed (see, e.g.,
(Refs. 14, 16, 17)).

\vspace{5mm}\centerline{\psfig{file=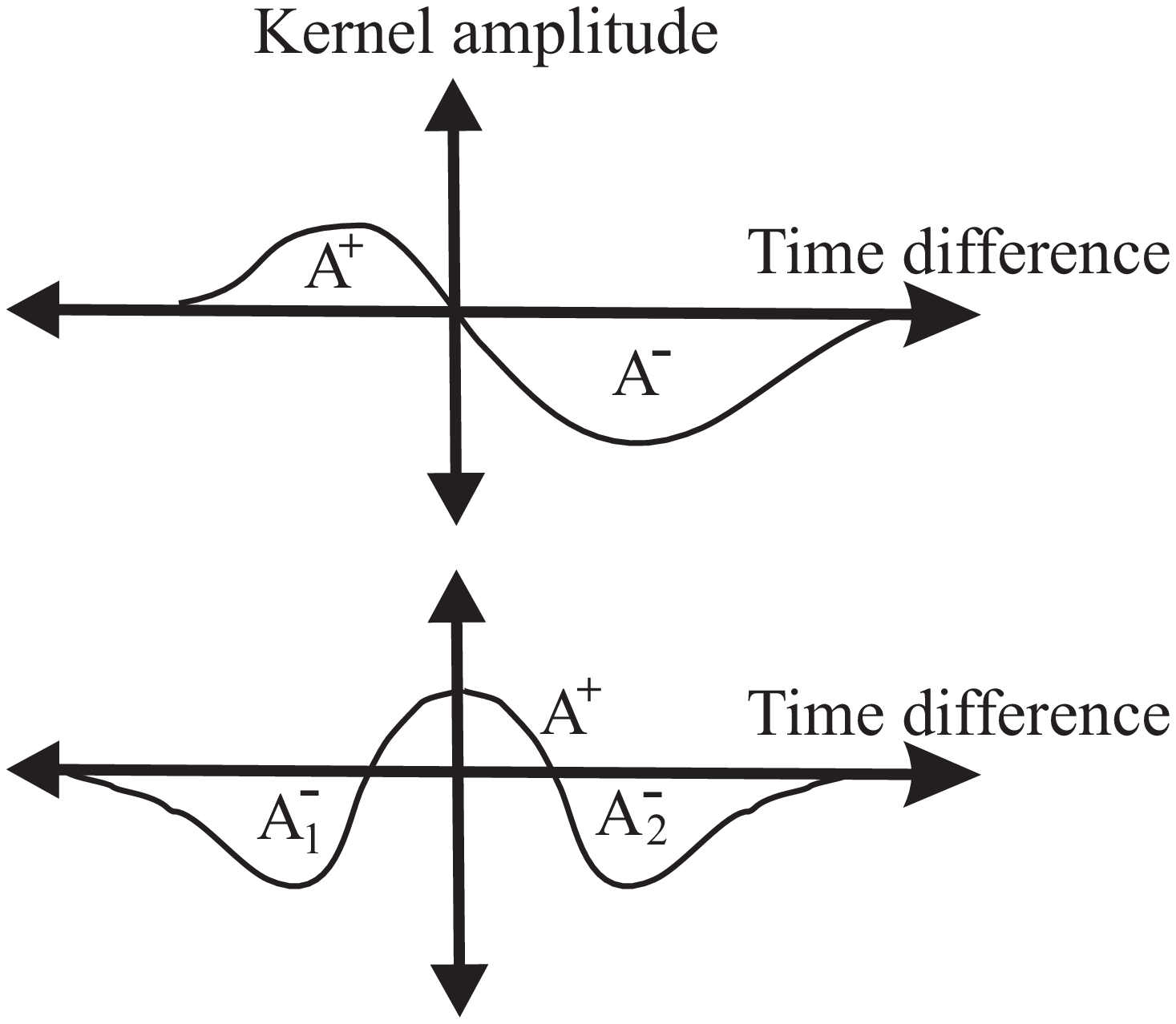,width=6cm}}
\vspace*{13pt} \fcaption {\textbf{Kernel
functions}\label{f:kernelf}} {\small Two temporal kernels as a
function of time difference between spiking time of neuron $i$ and
$j$ ($t_i-t_j$). Relevant parameter of the shape for
noise-sustained systems is the ratio ($r_{A^{+}/A^{-}}$) of the
areas (sums of positive  and negative parts/components) of the
kernel, $A^{+}$ and $A^{-}$, respectively
($r_{A^{+}/A^{-}}=A^{+}/A^{-}$).
  }
\vspace*{13pt}

In the first place, we have been interested in the emerging local
and global connectivity structure of $\mathbf{W}$. As the network
of the connections can be best described by a weighted graph, from
now on `nodes' stand for the neurons, while `edges' or `directed
edges' denote the connections among them. An insightful way of
characterizing graphs has been proposed by Watts and Strogatz.
They computed the characteristic path length ($L$), which is the
average number of edges on the shortest path in the network. They
also computed the clustering coefficient ($C$), which is large if
the average local connectivity is large. For more details, see
Ref. 18.

In this study, we applied the so called connectivity length
measure based on the concept of \textsl{network
efficiency}\cite{19}. This measure is more appropriate for
weighted networks\cite{20}, equally well applicable for describing
global and local properties and offers a unified theoretical
background to characterize our system. According to the
definition\cite{20,21}, \emph{local efficiency} between nodes $i$
and $j$ in a weighted network with connectivity matrix
$\mathbf{W}$ is $\epsilon_{ij}= 1/d_{ij}$, where $d_{ij}$
corresponds to the \textit{shortest path length} throughout all of
the possible paths from neuron j to i, where the path length
between each connected pair of vertices is the inverse of the
weight between them. For graphs with connection strengths of
values 0 or 1, $d_{ij}$ corresponds to the \textit{shortest
distance} between nodes $i$ and $j$. The average of these values
($E[d_{ij}] =\frac{1}{N(N-1)}\sum_{i\neq j}\epsilon_{ij}$)
characterizes the efficiency of the whole network. The local
harmonic mean \textit{distance} for node $i$ is defined as
\begin{equation}
D_h^l(i) = \frac{n^{(i)}(n^{(i)}-1)}{\sum_{j,k}\epsilon^i_{kj}}
\end{equation}
where $n^{(i)}$ is the number of neurons in subgraph $G^{(i)}$,
where subgraph $G^{(i)}$ consists of all nodes $l$ around neuron
$i$ with $w_{il}>0$, $\epsilon^i_{kj}$ is the inverse of
\textit{shortest distance} between nodes $k$ and $j$ in $G^{(i)}$.
$N>n^{(i)}$ arises when weights may become zero. In terms of
efficiency, the inverse of this value describes how good the local
communication is among the first neighbors of node $i$ with node
$i$ removed. That is why this measure can also be regarded as
\textit{local connectivity length}. It is a measure of the fault
tolerance of the system. The mean \textit{global distance} in the
network is defined by the following quantity:
\begin{equation}
D_h^g = \frac{N(N-1)}{\sum_{i\neq j}\epsilon_{ij}}.
\end{equation}
Global distance provides a measure for the \textit{size} (or the
diameter) of the network, which influences the average time of
information transfer. That is why, its inverse is used as the
(un-normalized) \textit{global efficiency}. According to the
literature\cite{20,21}, local harmonic mean distance measure
behaves like $1/C$ (inverse of the clustering coefficient),
whereas the global value is a good approximation of $L$ under
certain conditions.

Many different networks belong to the same structural family
regarded as `small-worlds'. Their most characteristic feature is
that they are efficient locally and globally, too. While local and
global connectedness are useful tools to characterize a network
architecture, it is worth investigating the degree distributions
of the incoming and outgoing connections as well\cite{22}. They
may provide information about the scaling of different properties
of the given structure, like the change of the diameter as a
function of the number of nodes. One particular subfamily of
small-world nets can also be characterized as `scale-free'
networks, because their most significant properties scale
according to power-law with the connection number distribution.
Most scale-free nets are also small-worlds, provided that
connection strength is not too sparse and basically no part of the
network is isolated.

\section{Results}
\label{s:results}

\vspace{5mm}\centerline{\psfig{file=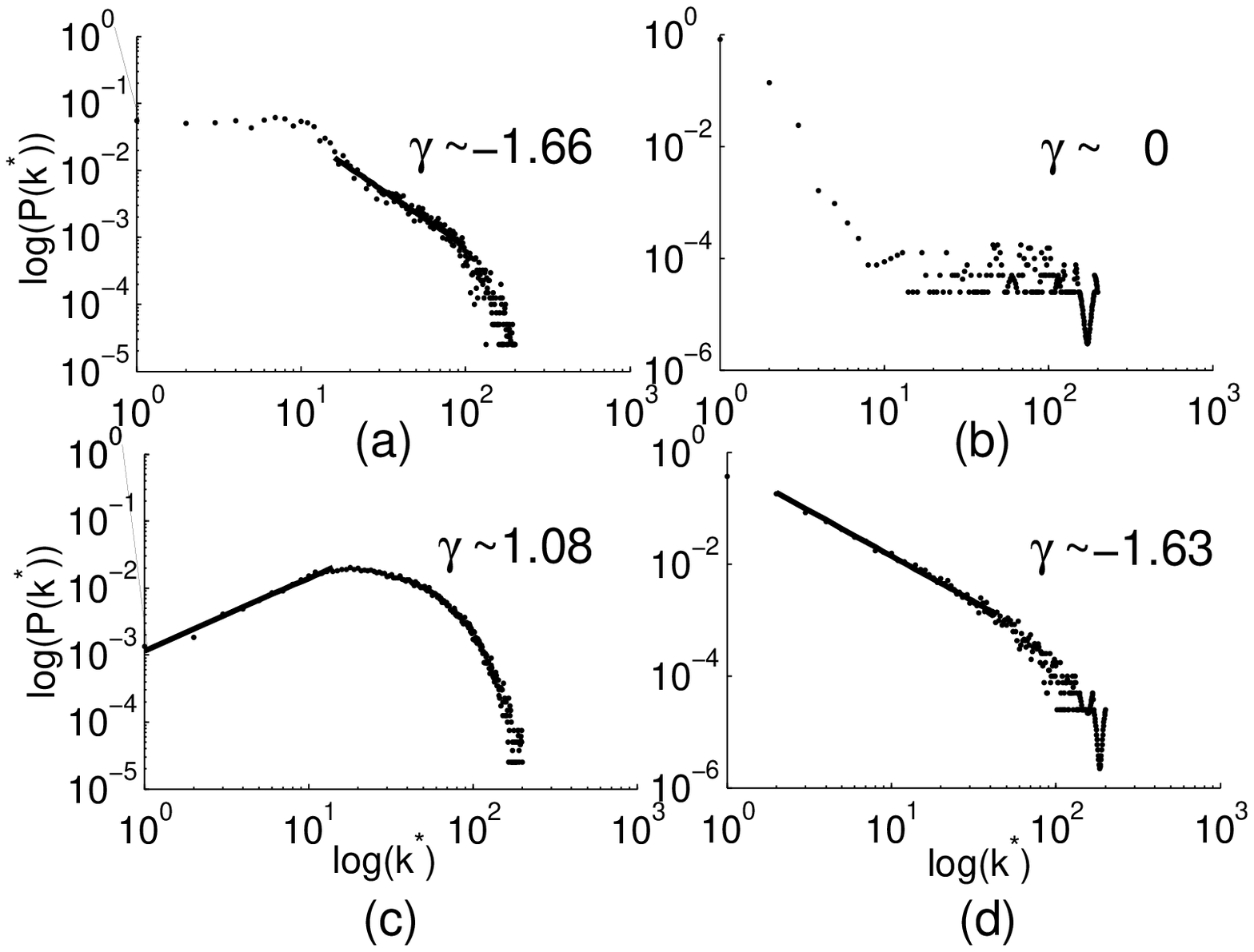,width=8cm}}
\vspace*{13pt} \fcaption{\textbf{Log-log plots for different
parameters}\label{f:loglog}}{\small
  The four diagrams display typical distributions for parameters (a): $r_{A^+/A^-}=0.1$ $r_{ex} =0.3$, (b):
  $r_{A^+/A^-}=0.1$
$r_{ex} =0.6$,(c): $r_{A^+/A^-}=0.6$ $r_{ex}=0.3$ and (d):
$r_{A^+/A^-}=0.6$ $r_{ex} =0.75$. Cases (a) and (d) are arbitrary
examples from the power law region. } \vspace*{13pt}

Figure \ref{f:loglog} summarizes our findings in different
parameter regions. The figure displays the emergence of scale free
nets as a function of the excitation level $r_{ex}$, the average
ratio of neurons receiving excitation from the environment, and
the ratio of the area of potentiation to the area of depression
($r_{A^+/A^-}$) in kernel $K$. The length of the scale-free
regions was determined by first plotting the distribution of the
sum of the weights of outgoing connections (averaged over 20 runs,
each run contains 10000 samples) for every parameter set studied.
Results are depicted on loglog plot. Supposing a power-law
distribution ($P(k^*)\approx k^{*\gamma}e^{-k^*/\xi}$, where $k^*$
denotes the discretized values of the connection strength), a
linear fitting was made to approximate $\gamma$. The width of the
scale-free region was estimated by the length of the region with
power-law distribution relative to the full length covered on the
log scale. Maximum error of the linear fit was set to $10^{-3}$
STD. That is, for 100 discretization points, the width of a region
spreading an order of magnitude on the loglog plot is equal to
0.5. Figure~\ref{f:matrices} shows the corresponding connection
matrices. While case (c) resembles a random structure, case (b)
seems to be a winners-take-most network, in which only a few
neurons dominate over the total amount of the connection strength.
However, cases (a) and (d) show strong clustering in a rather
sparse structure and therefore correspond to scale-free small
world networks characterized by their $\gamma$ values (~-1.66 and
-1.63, respectively). Figure.~\ref{f:matrices} depicts the
corresponding connectivity matrices.


\vspace{5mm} \centerline{\psfig{file=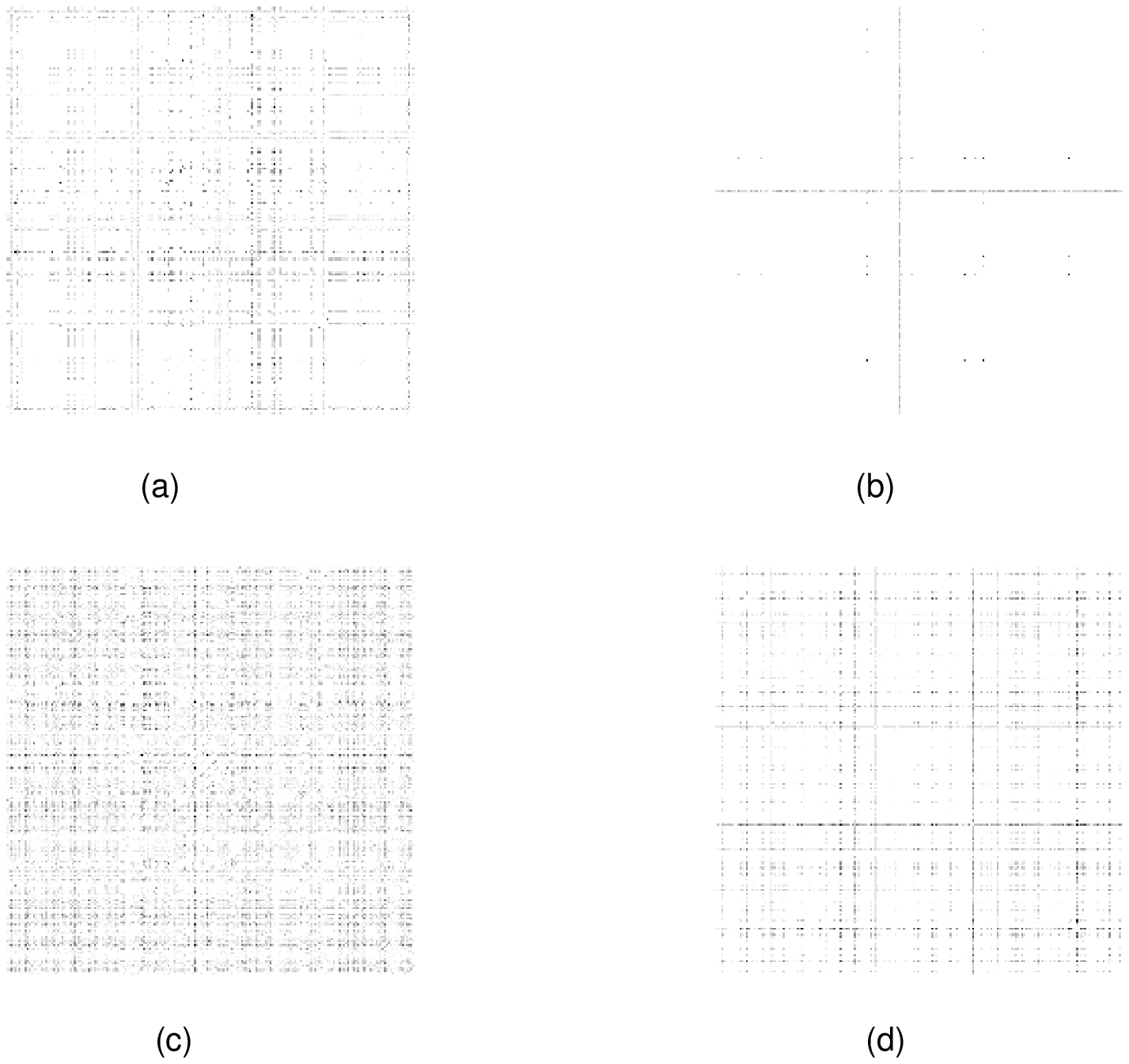,width=6cm}}
\vspace*{13pt} \fcaption{\textbf{Connectivity
matrices}\label{f:matrices}} {\small The four diagrams display
connectivity matrices corresponding to the cases in
Fig.~\ref{f:loglog}. Cases (a) and (d) are arbitrary examples from
the power law region.}

\vspace*{13pt}

With the help of the above introduced connectivity length measures
we studied also the emerging network structures as a function of
the following parameters: (i) the magnitude of the external
excitation and (ii) the strengthening--weakening area ratio
($r_{A^+/A^-}$) of kernel $K$. It can be seen that many connection
weights have been vanished and it has made possible to talk about
`subgraphs' with local connectivity. As an extreme case of the
general model, the binary neuron model was also investigated and
no important difference were found.

We compared the resulting HebbNet structures with a random net, in
which the same weights of the dynamic network have been randomly
assigned to different node pairs. Fig.~\ref{f:Harmdis} displays
the emerging connections of a HebbNet for two different parameter
sets. Figure \ref{f:Harmdis} highlights clearly the emerging
small-world properties, i.e., small local connectivity values
(high clustering coefficients) for case (d). Although the global
connectivity length was almost the same for all HebbNets and their
corresponding random nets, local distances are much smaller in
case (d). That is, connectivity structure is sparse but
information flow is still fault tolerant and efficient.

\vspace{5mm}
\centerline{\psfig{file=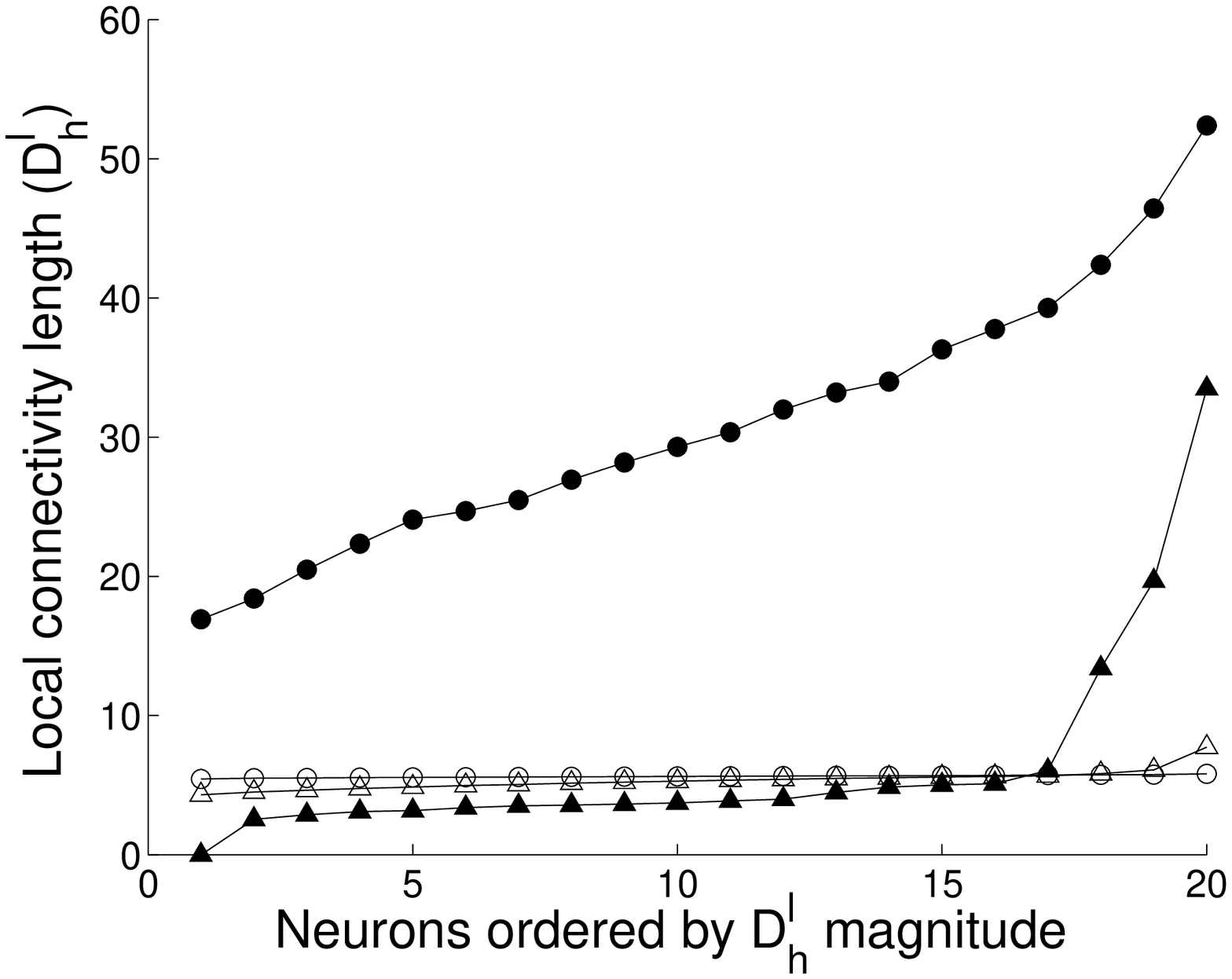,width=6cm}} 
\vspace*{13pt} \fcaption{\textbf{Local connectivity length
distances}\label{f:Harmdis}}{\small
  Local connectivity length distances in ascending order are shown.
  For better visualization not all data points are marked and the points are connected with
  a solid line. Lines with upward triangle markers: STDP learning. Lines with circles:
  same but randomly redistributed weights. Line with empty (solid) markers: HebbNet of
  case (c) (case (d)). Global harmonic mean distances for the original and
  for the randomized networks in case (c) of Fig.~\ref{f:matrices} (case (d) of Fig.~\ref{f:matrices})
  are about the same $D_h^g\approx D_h^{gr} \approx 5.5$ ($D_h^g \approx D_h^{gr} \approx 10$).
  }
\vspace*{13pt}

The robustness of the network to the external excitation (i.e.,
the amount of noise input to the network) is illustrated on
Fig.~\ref{f:robust}.

\vspace{5mm}\centerline{\psfig{file=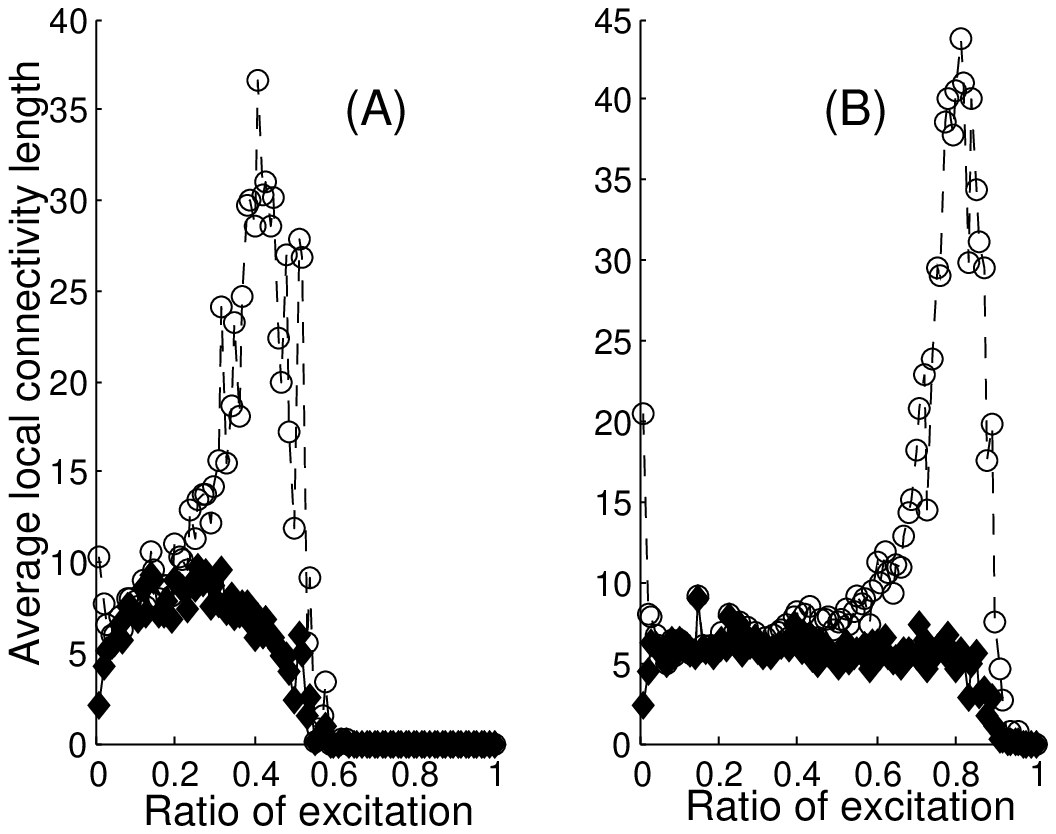,width=6cm}}
\vspace*{13pt} \fcaption{\textbf{Average local distance vs.
excitation ratio}\label{f:robust}}{\small
  A: $r_{A^+/A^-}=0.1$, B: $r_{A^+/A^-}=0.6$.
  Diamonds: average local distances for the evolving network.
  Circles: average local distances for the corresponding random net.
  }
\vspace*{13pt}

By increasing the excitation level, the average local connectivity
length of the random net is drastically increasing, whereas the
efficiency of the small-world network shows weak dependencies in
the same region. For the network with parameters $r_{A^+/A^-}=0.1$
(Fig.~\ref{f:robust}(A)), there is a sharp cut-off around
excitation level 0.55, where local distances suddenly drop, due to
the high ratio of excitation. Qualitatively similar behavior can
be seen for $r_{A^+/A^-}=0.6$ (Fig.~\ref{f:robust}(B)), but the
cut-off is around $r_{ex}=0.9$.

Results demonstrated so far characterize the `early' stages of
network development, as the interaction among neurons is weak due
to the low connection weight values in all of the above examples.
Figure \ref{f:feedback} demonstrates that even in case of strong
interaction, the found structural properties are present:
According to the figure, the power-law behavior is present in a
broad range of parameters. For the constant parameter of
Fig.~\ref{f:feedback} (i.e., for $r_{A^+/A^-}=0.1$) we have
experienced a convergence of the exponent of the power-law
distribution to -1.

\vspace{5mm}\centerline{\psfig{file=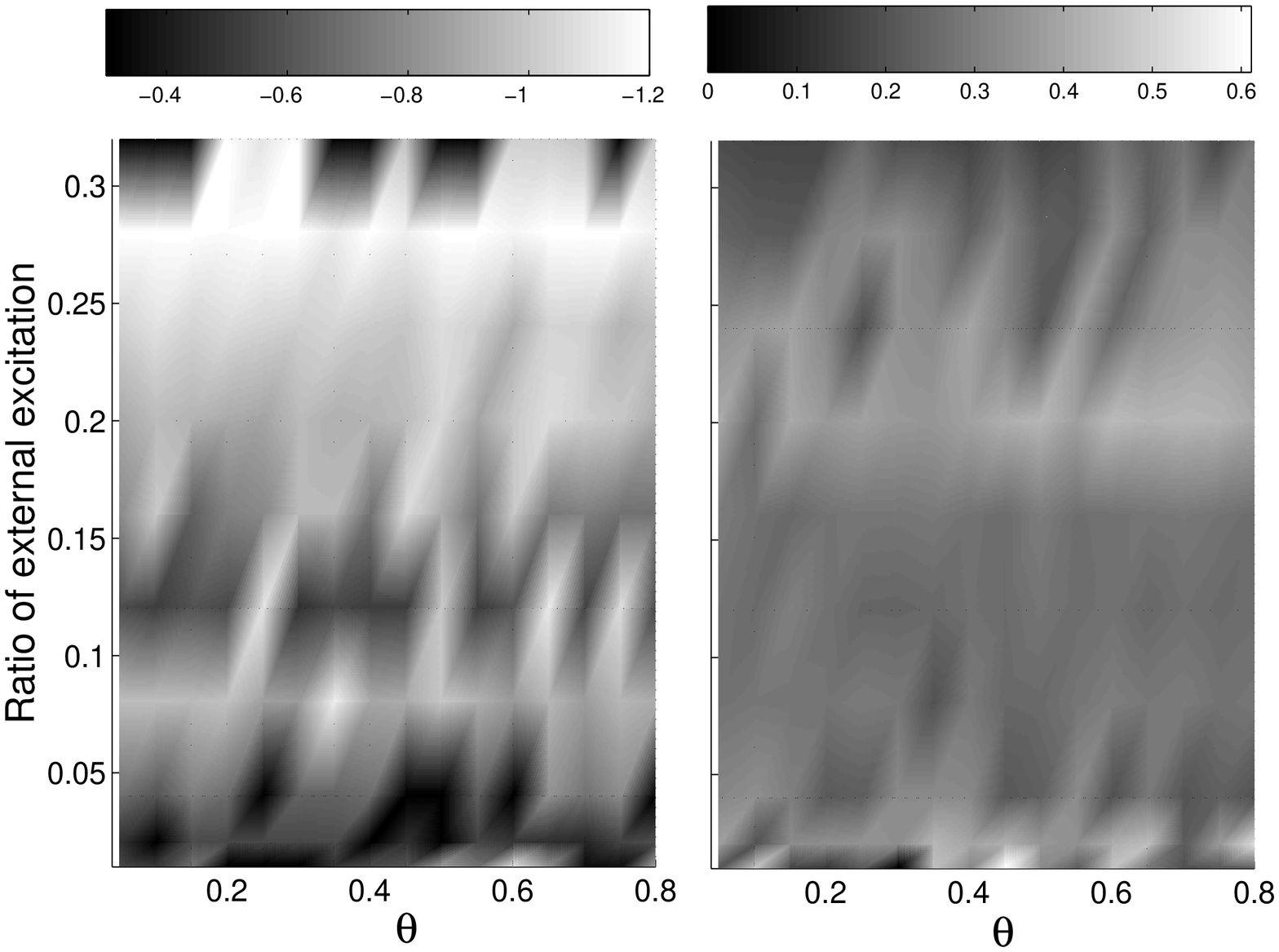,width=7cm}} 
\fcaption{\textbf{Power-law with significant
interaction}\label{f:feedback}}{\small
  \textbf{Left:} \textit{exponent of the power law},
  \textbf{right:} \textit{ratio of the power-law domain} (i.e., ratio of
  the width of power-law distribution region relative to the full length
  covered on the log scale) as a function of $r_{ex}$ and excitation threshold $\theta$.
  Parameter $r_{A^+/A^-}$ equals to $0.1$.
  Results are averaged over 700 steps. Input from other neurons could exceed
  the external inputs by a factor of 10. The power-law exponent is
  about -1 for broad regions of $\theta$ and $r_{ex}$.
  Outside these regions the network may vanish or start oscillating.
} \vspace*{10pt}

\section{Discussion and outlook}\label{s:disc}

One of the most exciting findings in recent scientific research is
that many complex interactive systems possess a surprising
structural and functional property: the emergence of scale-free
small-world networks (SFNs) of the building blocks. Such SFNs may
be found in distinct fields ranging from metabolic reaction chains
to social relation systems\cite{18,23,24,25,26,20,27}. One may
find SFNs in neurobiology as well. For example, the only case of
completely mapped neural network of the nematode worm C.
elegans\cite{29} is considered to form a small-world
network\cite{20}. An outstanding example is the Internet, which
displays this network structure at the hardware level of servers
and also at the level of web pages\cite{25,26,23}. This
fascinating self-organizing system has inspired several studies
and models. The original model of the the World Wide Web (WWW) by
Watts and Strogatz\cite{18} explored random restructuring of the
links among a finite number of `nodes'. Barab\'asi and his
colleagues introduced the concept of \textit{preferential
attachment} to model the WWW\cite{24,25}. The idea has been
extended to other types of networks\cite{26} and the focus has
been put on the search of general mechanisms underlying the
development of these distinct connection systems.

\subsection{Relation of HebbNet to other models}\label{ss:rel}
Although this paper is intended only to show some experimental
(simulation) results on noise induced network structures of
simplified neuron models, the results can be related to other,
well-known mechanisms, too. In the following we show that under
some (strong) constraining assumptions, our model can be
transformed to the model of Barabasi et al\cite{25}, the model of
preferential attachment. The following assumptions are made to
enable the above-mentioned transition:

\begin{romanlist}
    \item Let us suppose that at $t=0$ there are $N$ nodes, from which only
    $n$  nodes  ($n<<N$) have  at least one connection to other nodes.
    \item Let the changes in activity and connection strength be discrete
    by choosing both the weakening and strengthening step of the kernel to be
    of unit strength.
    \item Spikings of the cloud of ($N-n$) isolated nodes can be considered
    independent and the spiking probability is small. For such isolated nodes,
    only the external input, the second part of the right hand side of
    Eq. ~\ref{e:int_and_fire}, counts. Furthermore, the coincidence of spiking of two isolated neurons
    is negligibly small if the temporal kernel is short. At any time instant,
    when a neuron of the isolated cloud fires the nodes of the connected set
    may fire or not. If no coincidence occurs then there will be no change in the network.
    However, such coincidences are much more likely given the connectivity structure
    between the neurons of the connected set. This is so, because if one neuron fires
    then there is a chain of firing amongst these neurons.
    of th If they is In turn, the development of new connections between two
    isolated neurons is not likely, whereas isolated neurons tend to develop
    new connections toward the connected sub-net.
    \item In contrast to the cloud, the activity of the connected neurons
    is strongly dependent on the spiking activity of the `neighbors'. If firing starts
    in the connected cloud of neurons then the first term of the right hand side of
    Eq.~\ref{e:int_and_fire} will dominate the resulting firing chain. Input initiates
    the firing chain, whereas recurrent excitation from other nodes control that chain.
    In turn, the \textsl{probability} of firing can be taken as (approximately) proportional
    to the recurrent activity, controlled by the incoming connection
    distribution.
    \item Having established a connection between two nodes, it is
    kept steady and may not change by time. This is a strong assumption, which is
    tacitly assumed by the original model of preferential attachment, too.
\end{romanlist}

This latter constraint does not seem to be realistic in any model.
There is no reason that for a growing connection structure should
remain steady for old connections. Note, however, that random
rewiring of old connections can give rise to scale-free network
structure, too. In fact, this rewiring mechanism is the original
model of Watts and Strogatz\cite{18}. As it was noted at the very
beginning (see Section \ref{s:model}) our model has an intrinsic
weight redistributing property originated by the constraint that
$r_{A^{+}/A^{-}}<1$. In turn, the incremental growing of the
connected sub-net (by connecting new isolated neurons) and the
weight redistributing property of HebbNets can be seen as the
synthesis of the preferential attachment mechanism with continuous
new entries in the model of preferential attachment\cite{25}
\textit{and} the rewiring mechanism of the model of Watts and
Strogatz\cite{18}. That is, constraining our model lead to a
combination of two models both generating small-world structures.
Nonetheless, by means of numerical simulations we have shown that
our model can produce such connection structures  without the
explicit requirement on growing, and without a direct mechanism of
weight rewiring.

\subsection{Remarks on evolutionary systems}\label{ss:eval}
Interestingly, all the listed examples, one way or the other,
usually are also regarded as evolutionary systems. In our
particular case, the obtained results can also be interpreted in
an evolutionary context by reconsidering Edelman's alternative
neuronal group selection theory\cite{30} about the fundamental
role of selection during and after development of the nervous
system. According to Edelman, a theory to describe a system's
temporal change can be considered as `selectionist', if it
includes the following components:
\begin{romanlist}
  \item source of diversification leading to
variants,
  \item a means for encounter with an environment not initially
  categorized,
  \item a means for differential amplification over
some period of time of those variants in a population that have
greater adaptive value.
\end{romanlist}
It is no surprise that a system with these features falls into the
class of evolutionary systems as far as we look at the system as a
whole. In the nervous systems, there are at least two types of
temporal changes serving the first requirement: Diversification
can occur via the emergence of redundant connectivity during
development \textbf{and} via the modification of synaptic efficacy
during life-time learning. The main thesis of this paper is to
demonstrate how diversification can be realized by noise under
STDP rules. The second requirement is fulfilled if the pool of the
not yet seen input patterns is not
limited.\footnote{Considerations about the third requirement are
beyond the scope of the present study.}

Now, we can argue that noise in the nervous system has an
important role: Noise has no spatiotemporal structure. Thus,
obviously it cannot induce `learning' in general sense. However,
noise with STDP ---  according to in our results --- gives rise to
a search mechanism, which scans at all scales simultaneously.
Search in a scale-free manner can be most efficient if no
structural formation is known in advance. The searching feature is
robust: The noise generated structure is changing rapidly; results
depicted in the figures are averaged over several runs. The
continuous change induced by noise can be interpreted in the
following way. The noise \textbf{together} with the proportionally
expressed LTD and LTP mechanisms yields a continuous
sparsification and regeneration of the connections. LTP `chooses'
sound patterns, whereas LTD helps to `forget' those patterns and
maintains a competition amongst patterns. Synchronous patterns or
pattern series are quickly learned by HebbNets and approximately
stable connectivity patterns may emerge. Noise, in this case, may
modify the connectivity strengths and search may be performed
`around' an average stable connectivity pattern. Also, the noise
may help the system to escape from local minima. Noisy Hebbian
learning, in turn, is able to simultaneously learn correlations
and make selections among the discovered structures or patterns.

As far as other evolving networks are considered, the profound
implication of our result is that \emph{local} (Hebbian) learning
rules may be sufficient to form and maintain an efficient network
in terms of information flow. This feature differs from existing
models, such as the model on preferential attachment\cite{25}, the
global optimization scheme\cite{28}, and also from the original
Watts and Strogatz model\cite{18}.

In summary, we have demonstrated that small-world architecture
with scale-free domains may emerge in sustained networks under
STDP Hebbian learning rule without any other specific constraint
on the evolution of the net. According to our results, evolution
and plasticity of neural networks may be maintained by noise
randomly generated within the central nervous system. We
conjecture that the sustained nature of noise and the competition
imposed by appropriate $r_{A^+/A^-}$ values are the two relevant
components of plasticity and learning. It might be equally
important that exponents of HebbNets of neurons with significant
interaction are similar in a broad range of parameters providing a
system more stable against homeostatic parameter perturbations.

\section{Acknowledgements}
\vspace{0.5cm} This work was partially supported by the Hungarian
National Science Foundation, under Grant No. OTKA 32487.


\end{multicols}
\end{document}